\documentclass[english,american]{article}
\usepackage{mathptmx}

\usepackage[T1]{fontenc}
\usepackage[latin9]{inputenc}
\usepackage{listings}
\usepackage[letterpaper]{geometry}
\usepackage{babel}
\usepackage{latexsym}
\usepackage{amsmath}
\usepackage{makeidx}
\makeindex
\usepackage{graphicx}
\usepackage{nomencl}
\providecommand{\printnomenclature}{\printglossary}
\providecommand{\makenomenclature}{\makeglossary}
\makenomenclature
\usepackage[unicode=true,pdfusetitle,
 bookmarks=true,bookmarksnumbered=false,bookmarksopen=false,
 breaklinks=false,pdfborder={0 0 1},backref=false,colorlinks=false]
 {hyperref}

\makeatletter

\newcommand{\noun}[1]{\textsc{#1}}
\providecommand{\tabularnewline}{\\}

\usepackage{color}
\usepackage{xcolor}
\usepackage{listings}
\usepackage[font=small,format=plain,labelfont=bf,up,textfont=it,up]{caption}
\DeclareCaptionFont{white}{\color{white}}
\DeclareCaptionFormat{listing}{\colorbox{gray}{\parbox{0.99\textwidth}{#1#2#3}}}
\AtBeginDocument{
  
}

\makeatother

\addto\captionsamerican{}
\addto\captionsamerican{}
\addto\captionsenglish{}
\addto\captionsenglish{}

\begin{document}

\title{T2Script Programming Language}

\author{Piotr J. Puczynski {[}piotr@taboretsoft.com{]}\\
\\
T2Script Research Group at TaboretSoft\\
Department of Informatics and Mathematical Modeling\\
Technical University of Denmark\\
Richard Petersens Plads, build. 321\\
DK-2800 Lyngby, Denmark}
\maketitle
\begin{abstract}
Event-driven programming is used in many fields of modern Computer
Science. In event-driven programming languages user interacts with
a program by triggering the events. We propose a new approach that
we denote command\nobreakdash-event driven programming in which the
user interacts with a program by means of events and commands. We
describe a new programming language, T2Script, which is based on command\nobreakdash-event
driven paradigm. T2Script has been already implemented and used in
one of industrial products. We describe the rationale, basic concepts
and advanced programming techniques of new T2Script language. We evaluate
the new language and show what advantages and limitations it has.
\end{abstract}
\begin{center}
\includegraphics[width=0.7\textwidth]{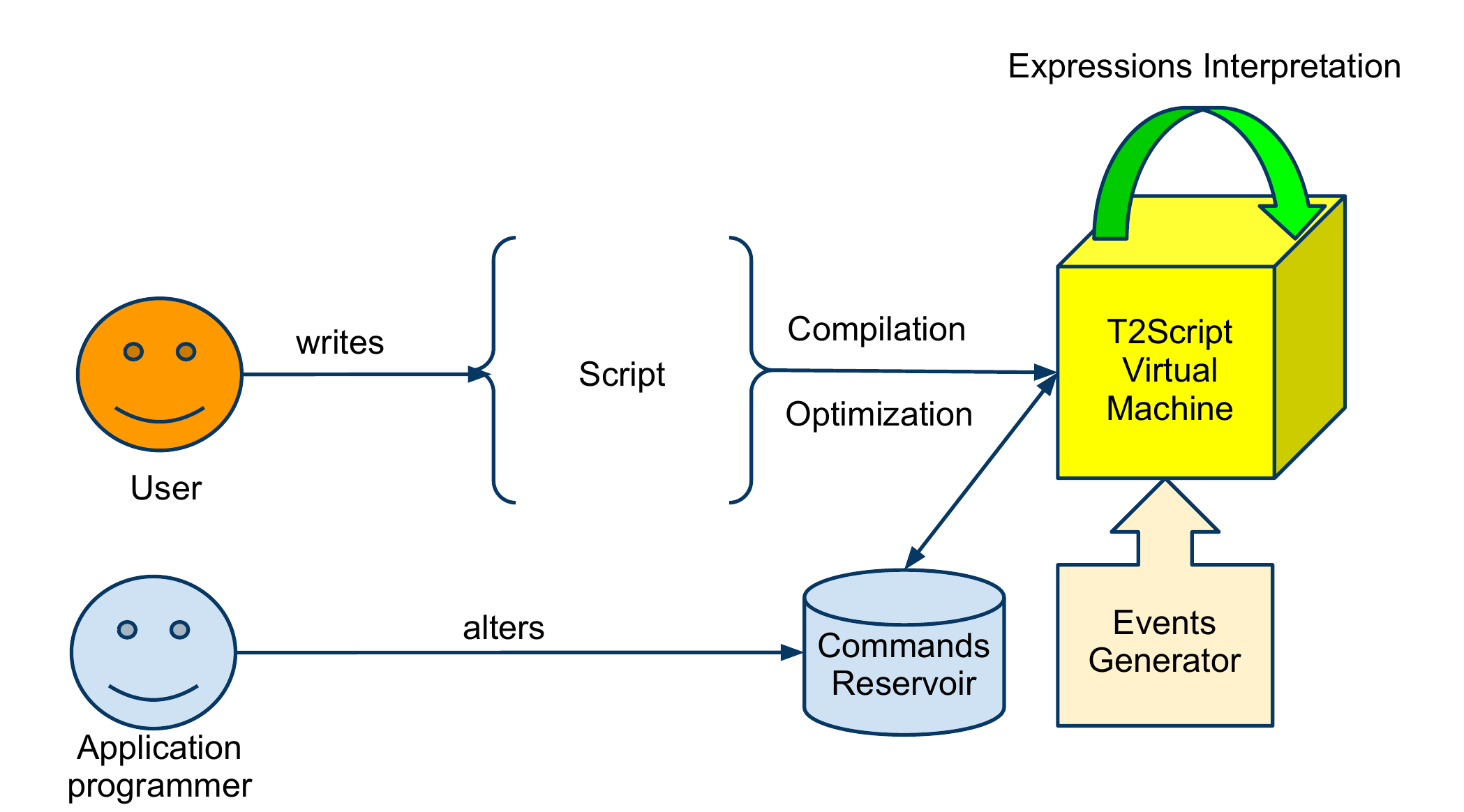}
\par\end{center}

\pagebreak{}

\tableofcontents{}

\pagebreak{}

\part{Introduction}

\section{Foreword}

This is a vision document describing fundamental concepts in a new
dynamic language \emph{T2Script}. The language, including all former
and current versions, has been designed by Piotr J. Puczynski in years
2003-2010. The language is \emph{command-event driven} and designed
to allow fast purpose-specific operations for handling strings. It
is also designed to be integrated with other programming languages
by means of metaprogramming techniques. The language can be used for
many applications in both academia and industry.

We would kindly like to thank Hubert Baumeister and Michal Staszewski
for reading the first versions of the manuscript and giving a feedback.
Also we would like to thank the members of community of Taboret2 --
that sometimes acted as guinea pigs for the subsequent language versions
-- for important feedback during the language design and for pointing
out problems and possible improvements.

\section{Motivation\label{sec:Motivation}}

The event-driven programming languages are usually used to support
\emph{Graphical User Interfaces} (GUI) \cite{Ferg2006}, network and
sometimes to implement robust or real-time systems \cite{Dabek2002,Ghosal2004}.
In this paper, we will show a new language, T2Script, which constitutes
event-driven paradigm for purpose-specific language. The event-driven
programming environments have many applications in telecommunications
and \emph{Human-Computer Interaction} (HID) systems. In these dynamic
environments, the program needs to handle many events in a short time.
Our T2Script language aims to solve the problems and issues:
\begin{enumerate}
\item In events-driven languages the interaction with the user is limited
to the events. We would like to increase the scale of interaction
in our language by introducing \emph{commands }(as element of language).
This approach is very useful in many of specific applications that
the language could be adapted for. We will denote that approach as
\emph{command-event driven }(CED\emph{\nomenclature{CED Programming}{Command-Event-Driven Programming. A new paradigm of programming that is proposed on base of T2Script programming language. It introduces user's interactions with events and commands.}})\emph{
programming paradigm};
\item In most of the languages like \emph{Java,} the events dispatcher is
exposed to the \emph{user} (i.e. script programmer) as listener that
is not an internal part of language itself. For simple applications,
this approach may be seen as too complex for a given problem and may
discourage a user to follow it.
\item There is no event-driven specific-purpose language that was designed
for strings handling in which the language expressions are being separated
from the text, not text being separated from the expressions (that
essentially makes a difference of importance);
\item The existing languages don't allow easy extensions with modules written
in other programming languages. And if they do the technique usually
increases program execution time;
\item Problem of static code analysis for security, particularly, for contract
enforcement;
\item Standard programming languages are difficult (if not impossible) to
improve and change.
\end{enumerate}
We will show our solution generically but we will also mention the
example implementation of T2Script language that was build-in as a
module to \emph{Taboret2 }application that is popular chat software
in Poland (see Section \ref{sec:Taboret2-Application}). This application
extends the commands set of T2Script language with \emph{Internet
Relay Chat} (IRC) client-server specific interactions.

\subsection{Contribution and novelty}

Few significant contributions are made by this paper. It is the first
publication about the new dynamic programming language: T2Script.
The document is a description revealing language elements and presenting
how to use them from the point of view of the user (and, in limited
scope, application programmer). It shows how the problems presented
in Section \ref{sec:Motivation} being solved in the language design
and build-in language commands. Document also contains description
of CED Programming paradigm that is proposed based on T2Script. The
Section \ref{sec:Evaluation} contains the very first try to evaluate
the new language.

Note, this is the vision document and the problems are inspected from
high-level remoteness. The detailed descriptions of solutions of some
problems are left for future work.

\section{Taboret2 application\label{sec:Taboret2-Application}}

Taboret2%
\footnote{Taboret2 website: http://www.taboret2.pl%
}\nomenclature{Taboret2}{Industial application in which T2Script was implemented for the first time. It is customized IRC client. http://www.taboret2.pl/en}
is chat software popular in Poland. It is designed to be the client
application of \emph{Onet.pl} extended IRC servers -- the biggest
Polish Internet portal and 3rd most accessed website in Poland (according
to Alexa.com data from January 2011). Taboret2 was founded in 2002
and is constantly developed since.

The previous versions of T2Script language that we describe had been
used in the application and had number of developers in real-world.
Mostly due to program localization issues (Polish interface of Taboret2)
and absence of good documentation the language is not known to the
public. This document aims for a change in this subject.

\section{Language history and development\label{sec:Language-History}}

\begin{figure}[h]
\begin{centering}
\includegraphics[width=0.9\textwidth]{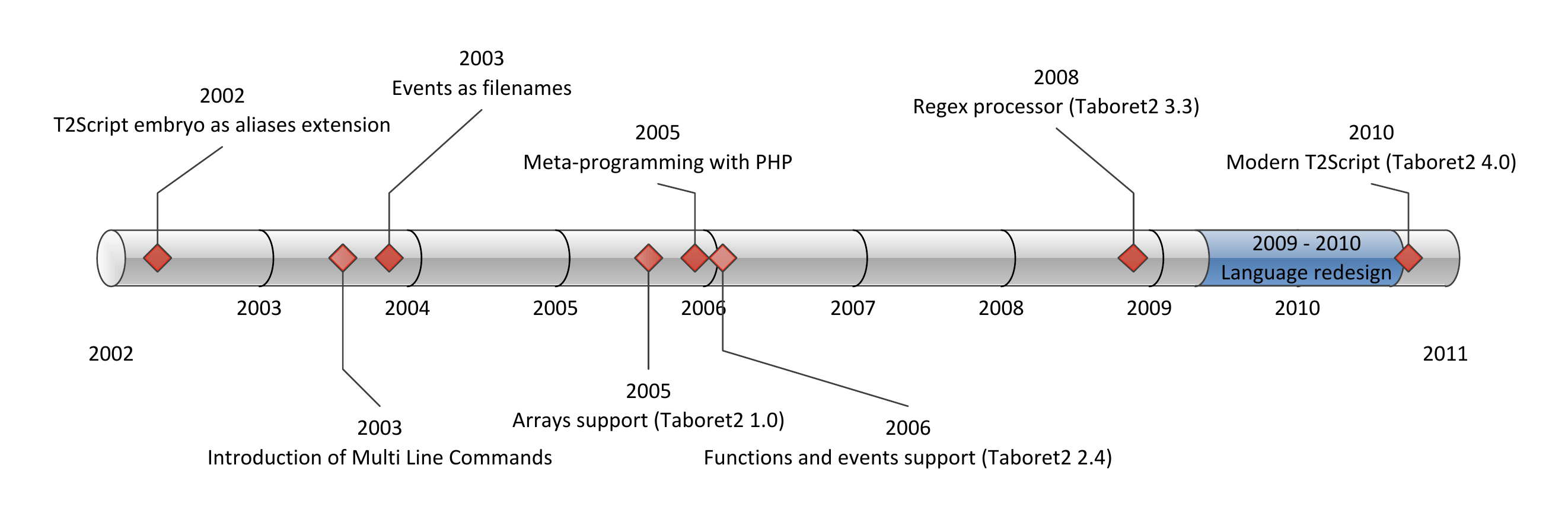}
\par\end{centering}

\caption{Timeline of Evolution of T2Script.\label{fig:History-of-Evolution}}

\end{figure}

The current version of T2Script was introduced for the first time
in the software bundle \emph{Taboret2 4.0} in the end of 2010\emph{.}
For simplicity, we will assign the language version to be \emph{4.0}.
Providentially, this assignment also corresponds to the language evolution
as shown in fig. \ref{fig:History-of-Evolution}.%

T2Script notation comes from {}``Taboret2 Script''. This name was
chosen at the beginning of language development as at that time T2Script
was an internal inseparable Taboret2's part. T2Script was born in
2002 when first extension to aliases\index{alias} was developed in
the application. Aliases were simple user customized commands redefinitions.
The aliases definition required more complex commands handling in
the engine. 

Subsequently, users required more complex expressions for more complex
aliases. All this led to introduction of \emph{Multi Line Commands}
(MLC\nomenclature{MLC}{Multi Line Command. The historical command that was used to enable to execute multiple commands in one line.})
that encapsulated many language commands in one line (so the user
would be able to input it in text field). The complexity of usage
of MLCs was later solved by introduction of {}``scripts'' saved
in files. At that time (2003) also first events support (recognized
on the base of filenames) was added. Functions support was added later
in 2006 and solved the file names based events limitations.

Before language redesign in 2009-2010 it was not possible to reuse
the language apart from Taboret2. The mentioned language redesign
was a breakthrough in T2Script history. Language was now designed
as a separable module of Taboret2 with possible future extension to
other applications. MLCs (that were still used for some purposes before)
became obsolete and the language became more expressive. Many advanced
mechanisms (like exceptions handling) were added to the language.

In this document, we will describe the modern version of T2Script
(T2Script 4.0) that crystallized after the 2009-2010 language redesign.

\part{Basics}

\section{Language purpose}

\begin{figure}[h]
\begin{centering}
\includegraphics[width=0.7\textwidth]{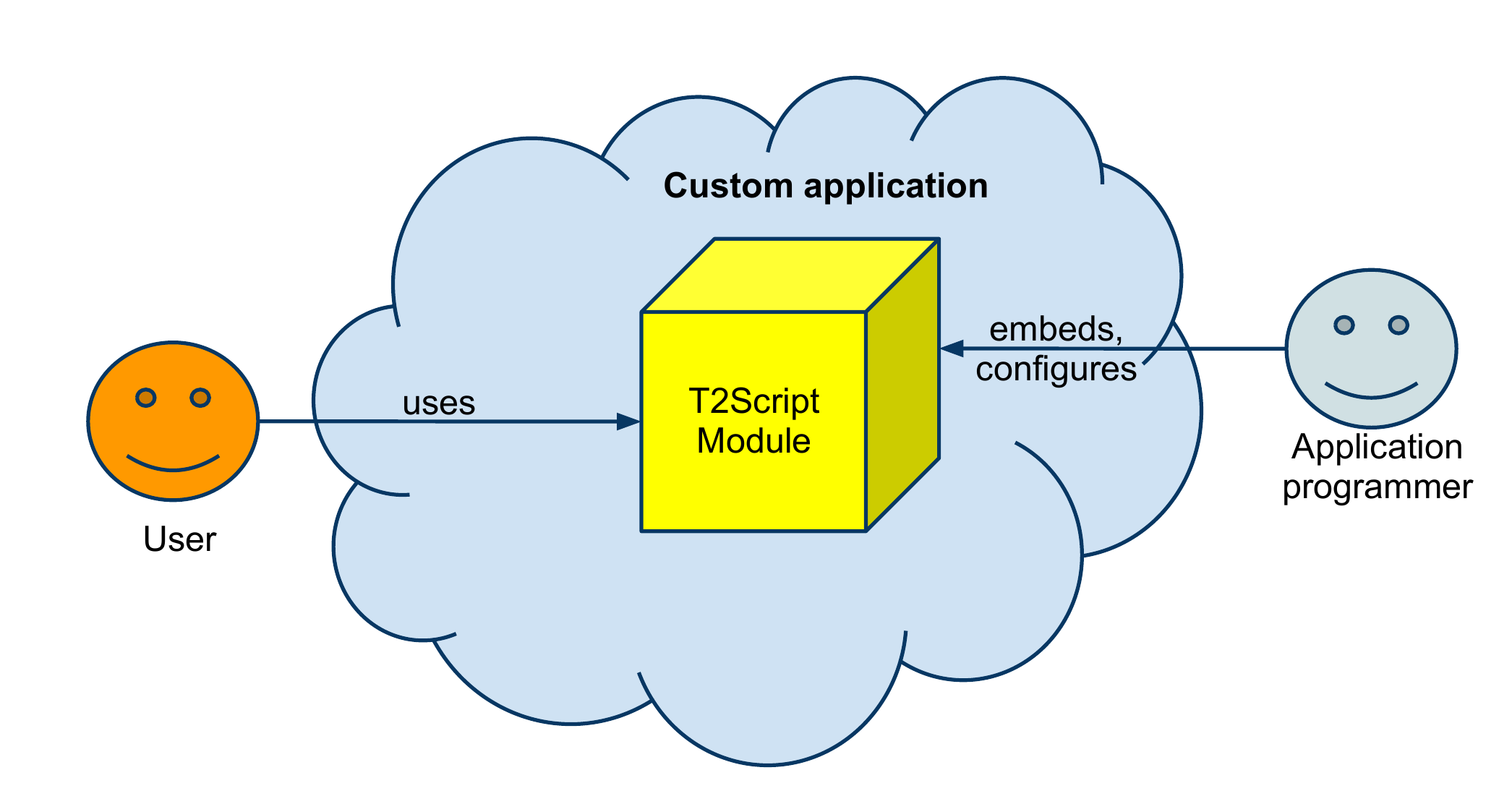}
\par\end{centering}

\caption{T2Script usage in custom application.\label{fig:T2Script-usage-in}
Application programmer first embeds and configures T2Script Module
in her application, then the module is used by a user.}

\end{figure}

\nomenclature{User}{By user of T2Script language we define the script programmer that uses the application with configured T2Script module (script level) to build scripts and use commands.}\nomenclature{Application programmer}{Application programmer embeds and configures the module of T2Script in her application (application level).}The
purpose of T2Script programming language is to enable the \emph{custom
application} (later denoted also as \emph{application}) to use commands
and events interfaces. The language module is embedded and configured
in custom application by \emph{application programmer}, then used
by \emph{user} (see fig. \ref{fig:T2Script-usage-in}). %

Because commands nature is text-based, the T2Script language is mainly
designed for handling strings. The idea of designing embeddable command
programming language is not new. In 1989 John K. Ousterhout designed
\emph{Tool command language} (Tcl\index{Tcl}) \cite{Ousterhout1989}.
Tcl can be reused in many software applications as suggested in \cite{Sametinger1997}.
In Tcl an event can be bind to application specific command. T2Script
allows user to define and trigger own events, as well. Tcl is implemented
in C that may not be suitable for nowadays applications without using
additional libraries. T2Script is implemented in C++ and supports
object-oriented structure when embedded in application. Nevertheless,
indisputably there are many similarities in the language purposes
shared between Tcl and T2Script.

T2Script comprises following design objectives:
\begin{itemize}
\item The language principal entities are \emph{command}s\index{command}
and \emph{event}s\index{event}. Commands and events are divided in
3 groups: language internal, application specific and user defined.
The purpose of commands is to enable users of application to run them.
The events are defined as actions that occur at specific circumstances
in the application.
\item The language is programmable. Application programmer and application
itself are able to change the commands set in run-time. Users and
application programmers can also define new events and trigger them.
\item For simple usage, the language is transparent for the user. I.e. for
simple commands the user may not even be aware of using the programming
language structures.
\item The language is efficient for interpretation at run-time. It is always
a trade-off between programming language expressiveness and simplicity
of machine interpretation. The dilemma was mentioned in the book by
John C. Mitchell that belongs to classic literature of programming
languages design \cite{Mitchell2003}. In case of T2Script it is important
for the language expressions to be interpreted with low cost as the
scripts are only precompiled and run in virtual machine. Events can
also occur very often subsequently or parallelly.
\item The language provides a simple structured interface to be adopted
in many existing programs. The time needed for embedding the language
into custom application should not exceed 24 hours of work of advanced
programmer.
\item The language allows low-cost metaprogramming techniques. This approach
makes it easy to write the programs using T2Script commands in other
programming languages and execute them in T2Script environment.
\end{itemize}

\section{Scripts interpretation process}

\begin{figure}[h]
\begin{centering}
\includegraphics[width=0.7\textwidth]{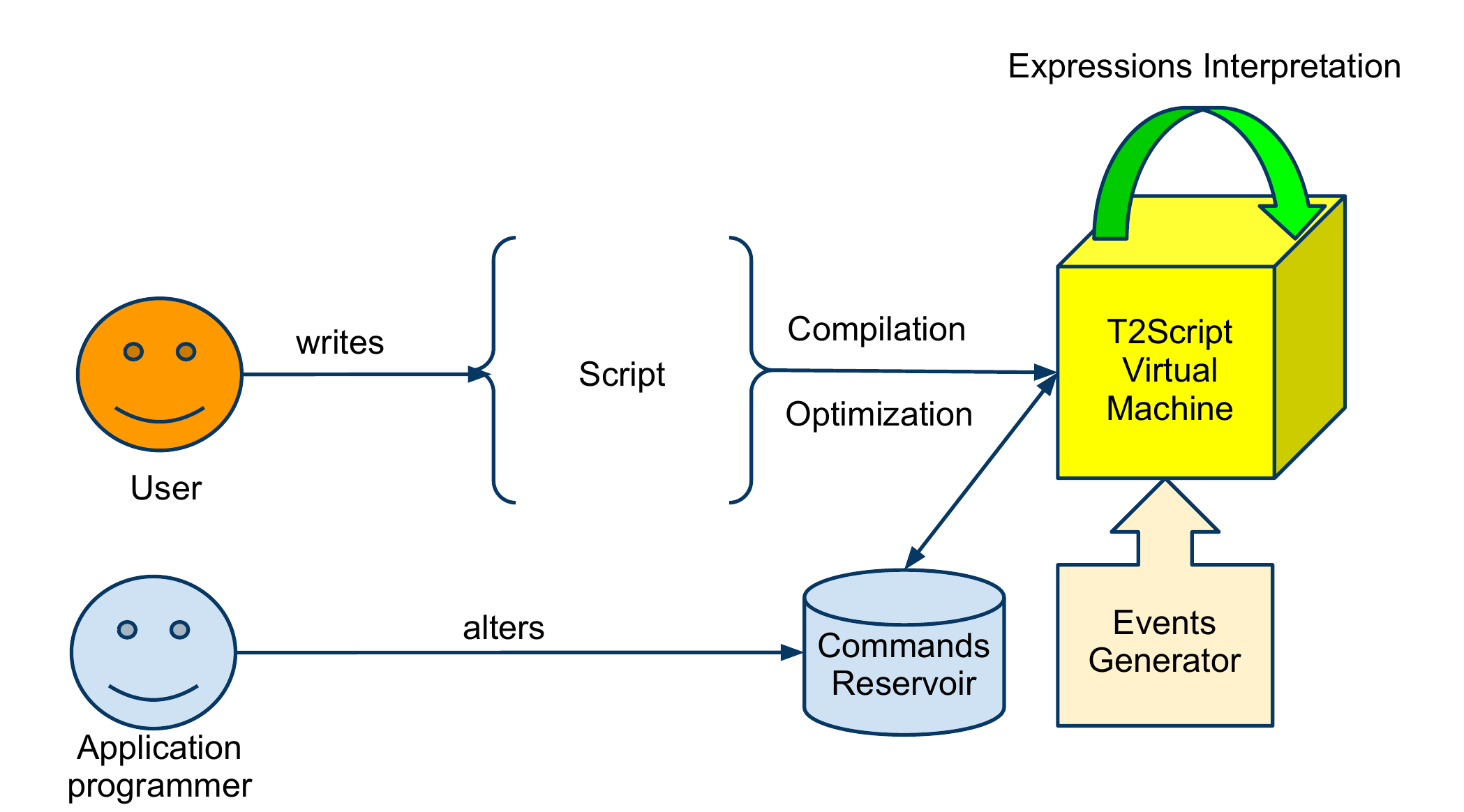}
\par\end{centering}

\caption{User and application programmer interactions with T2Script implementation
elements. Note, interacts with T2Script engine by supplying script
(single command or script file) and application programmer interacts
by defining and altering an entries in Commands Reservoir.\label{fig:User-interactions-with}}

\end{figure}

Scripts in T2Script are translated to internal bytecode format that
is in \emph{directly interpretable representation} (DIR) form before
execution on T2Script application-level process \emph{Virtual Machine}
(VM) \cite{Smith2005,Rau1978}. The VM \emph{Application Programming
Interface} (API) is the current set of commands (\emph{Commands Reservoir\index{Commands Reservoir@\emph{Commands Reservoir}}})
that can be altered by the application programmer (or, at run-time,
by application). The general scenario of scripts running on T2Script
VM is shown in fig. \ref{fig:User-interactions-with}. %

User supplies T2Script file in one of two forms:
\begin{enumerate}
\item Single-command\index{command} script received from user's input (e.g.
from the console);
\item Script file with functions and events definitions.
\end{enumerate}
These two are different for the T2Script VM. The single command is
compiled and executed immediately with specific \emph{command's context}.
The script file also contains the definitions of \emph{functions}
and \emph{events} that cannot be supplied with single-commands. It
is optimized during the translation to internal form by efficient
optimization techniques \cite{Arnold2005} and command caching. Next,
the translated DIR is loaded into VM (occurs only once). Then the
script is set to idle state and run on VM each time \emph{Event Generator}
triggers an event. Scripts files are executed using \emph{default
context} (\emph{context} concept is described in section \ref{sec:Commands}).

The script files in T2Script currently use format \noun{.tsc\index{tsc@\noun{tsc}}\nomenclature{TSC}{T2Script File Format. Format of distribution the script file containing source code, functions and events definitions.}}
and are distributed in source code. Each script file represents logically
a separate module. The translation to DIR is performed after the script
is loaded in the memory (e.g. at application startup or when user
calls \noun{load\index{load@\noun{load}}} command). T2Script fully
supports \emph{Unicode\index{Unicode@\emph{Unicode}}} and script
files can be saved using \emph{Unicode} standard \cite{Aliprand2004}.

During the run on T2Script VM all scripts are interpreted as they
still contain simple \emph{expressions} in the format described in
section \ref{sec:Expressions}. This process has the nethermost low-cost
because of existence of \emph{expressions preambles} and characteristic
discriminative expressions format.

\section{Commands\label{sec:Commands}}

\nomenclature{Command}{Command is basic T2Script element that has a name and optional parameters. Commands can be run and added by the user.}Commands\index{command}
are an underlying concept of T2Script. Commands are used as the language
keywords as well as to call application defined operations.

\subsection{Format of commands\label{sub:Format-of-commands}}

\begin{figure}[h]
\begin{centering}
\includegraphics[width=0.7\textwidth]{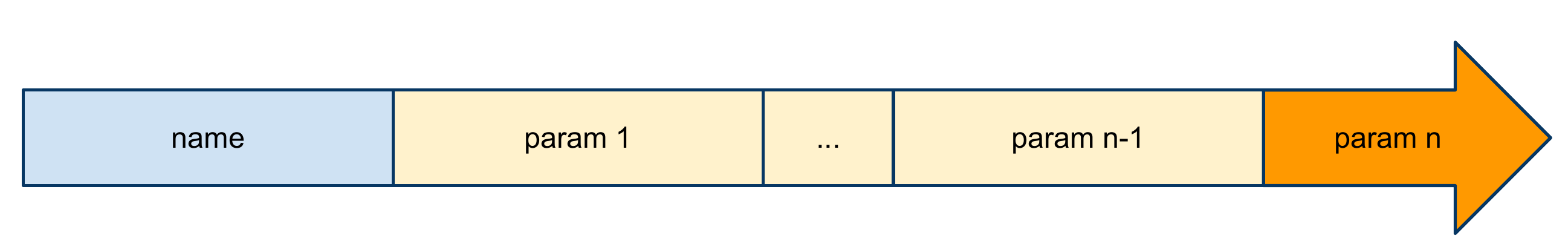}
\par\end{centering}

\caption{Structure of command accepting \emph{n} parameters in T2Script. Identical
colors of elements express similarity. The presented elements are
separated with at least one space. Consequently, only the last, arrow-shaped
element -- parameter \emph{n} -- is allowed to contain spaces.\label{fig:Schema-of-command}}

\end{figure}

The command is identified by its name. Name consists of arbitrary
number of Unicode characters except space, {}``\#'', {}``;'',
{}```'', {}``|'' and {}``/''. Command accepts from 0 to many
\emph{parameters}\index{parameters}\nomenclature{Parameter}{Parameter is an argument of command. Command can contain from 0 to infinite number of parameters. Parameters are separated by at least one space.}.
Fig. \ref{fig:Schema-of-command} presents the structure of command
in T2Script.%

Every parameter and a name are separated from each other by one or
more spaces. Only the last parameter\index{parameters} of the command
is allowed to contain spaces. The example of multi\nobreakdash-parameter
command is presented in listing \ref{lis:Example-multi-parameter-command.}.
Note, the command \noun{setvar\index{setvar@\noun{setvar}} }accepts
2 parameters and in the example, second parameter consists of several
words. The semicolon (line-ending character in file scripts) for single\nobreakdash-command
supplied from user input is skipped. In all the examples, we assume
the commands are placed in the script file, and therefore they must
have a semicolon\index{semicolon} after the last element.

\selectlanguage{english}%
\lstset{float=h, language=, basicstyle=\ttfamily, numbers=none, numberstyle={\tiny}, breaklines=true, frame=bottomline}

\selectlanguage{american}%

\begin{lstlisting}[caption={Example multi-parameter command.},{float=h},label={lis:Example-multi-parameter-command.}]
setvar my_var Hello, this is multi words parameter;
\end{lstlisting}

\begin{figure}[h]
\begin{centering}
\includegraphics[width=0.7\textwidth]{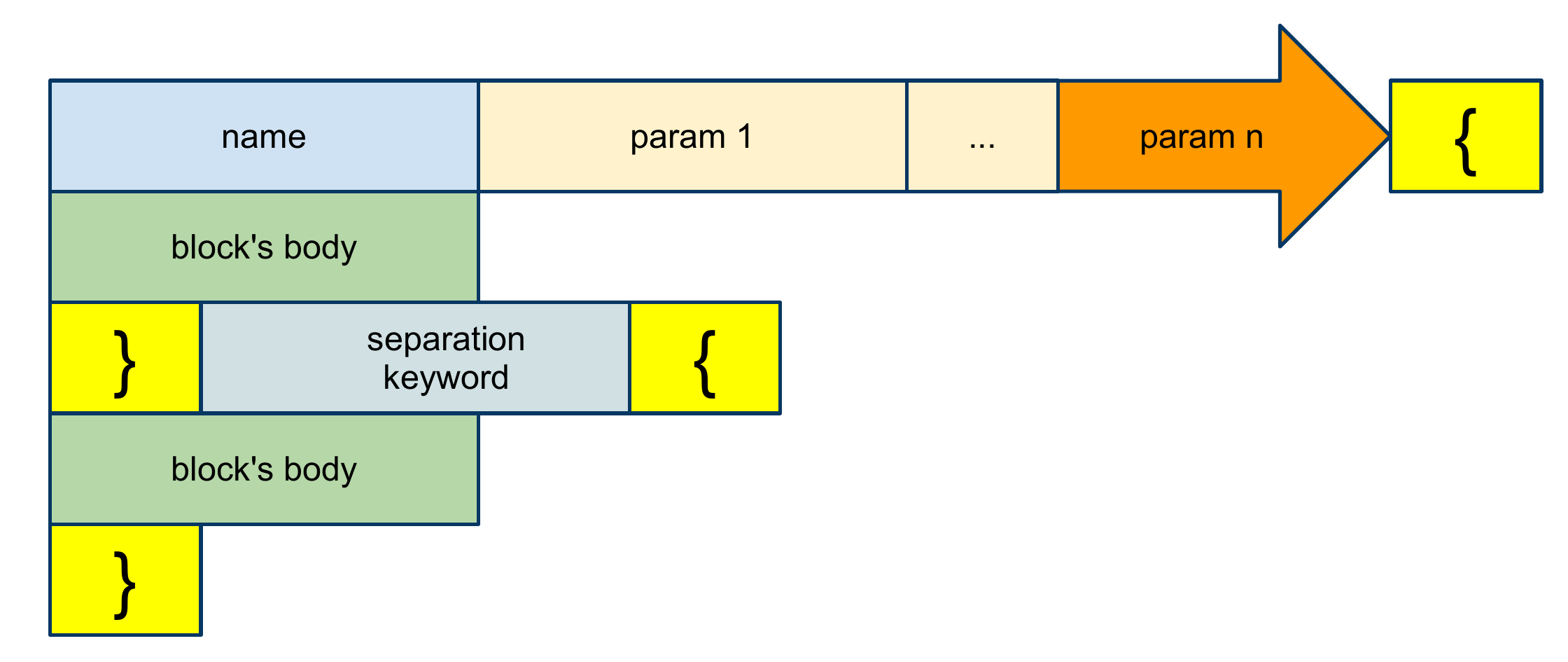}
\par\end{centering}

\caption{Structure of block-command accepting \emph{n} parameters and 2 blocks.
Identical colors of elements express similarity. The bodies contain
other commands. The second block with separation keyword can be set
as optional.\label{fig:Schema-of-block-command}}

\end{figure}

Second more complex and powerful command format in T2Script is block-command\index{block-command}
presented in fig. \ref{fig:Schema-of-block-command}.%
 Block-commands are used only in the script-files and they allow defining
more sophisticated commands like language keywords. The block-command
has optionally one or two blocks. The second block must have \emph{required}
or \emph{optional} attribute. Each block contains one or more commands
(including multi\nobreakdash-parameter and block\nobreakdash-commands).
The example \noun{if\index{if@\noun{if}}} condition command (that
is a block-command) in T2Script is presented in listing \ref{lis:Example-block-command.}\@.

\begin{lstlisting}[caption={Example block-command.},{float=h},label={lis:Example-block-command.}]
if $success {
	setvar @text Patient fine;
} else {
	setvar @text Patient still sick;
}
\end{lstlisting}
For the presented example, \noun{else} is separation keyword. The
keyword and the following block are optional.

\subsection{Case-sensitivity}

The names of the commands are case-insensitive. It is exception as
all other language's elements (including variables and function names)
that are case-sensitive. The rule stems from:
\begin{enumerate}
\item compilation process that effectively removes the names of the commands
from the bytecode and replaces them with identifiers that point to
VM commands functions;
\item the user input method when the language is used in \emph{interactive
mode} in the console (command name is traditionally case-insensitive
in the consoles).
\end{enumerate}

\subsection{Multi-line policy}

User can distribute command over many lines in the script file. Logical
line separator is semicolon\index{semicolon} ({}``;'').

In normal situation, when command is distributed over many lines the
space is added as a connecting character. This is avoided by adding
acute ({}```'') before connection (compare listing \ref{lis:Command-wo-connector}).

\begin{lstlisting}[caption={Command distributed over many lines with and without connector. },{float=h},label={lis:Command-wo-connector}]
// space connector is added
setvar multi My name is
	James;

// space connector is avoided
setvar Pi 3.1415926535`
	8979323846;
\end{lstlisting}

In the example, the value of \emph{multi} is {}``My name is James''.
Value of \emph{Pi} is {}``3.14159265358979323846''.

In the blocks, the semicolon preceding the closing element of the
block ({}``\}'') and following the last command in the block can
be skipped but it is not recommended. Semicolon after block-command
is skipped.

We also note the comments\index{comments} in T2Script are always
preceded by {}``//''. Additionally for respecting the comments,
{}``//'' must be the first non\nobreakdash-white characters in
the line.

\subsection{Context of commands\label{sub:Context-of-commands}}

Every command in T2Script contains a context\index{context}. Context
is background information of command execution. From the point of
view of application programmer, context is an object that she defines
\emph{a priori} and attaches to the command. Context is then used
to decide which commands and constants are available in the environment.
For instance, in Taboret2, implementation context is IRC channel object
and it sets some specific commands and constants for each window in
the application in the way the user that runs the same command in
different windows will get different results. Some specific commands
and constants are only available for some windows.

Context can be changed at run-time by application programmer for single-commands.
Dynamic changing of context functionality is not part of the language
but can be exposed to the user. E.g. in Taboret2 application, this
functionality is exposed to the user in the command \noun{wnd} that
changes the window of command context.

It is optional to set the context manually. If command context is
not set, command executes with default context. That is the case of
commands in the text scripts from the functions.

\subsection{Result of command}

From the point of view of application programmer, every command returns
a result. The result is a Boolean value. If the value is \emph{false}
and the command error field is set then it is considered that the
error occurred during the command execution. If the value is \emph{false}
and the value of error is not set or is set to specific constants
defined in T2Script VM it has other meaning for definition of more
complex language structures like \noun{break\index{break@\noun{break}}}
mechanism in loops. Table \ref{tab:The-error-codes} presents the
existing commands results and error codes combination and their effects
for languages structures build into T2Script.%
\begin{table}[h]
\begin{centering}
\begin{tabular}{|c|c|c|}
\hline 
Result & Error code & Effect\tabularnewline
\hline 
\hline 
true & - & command successfully executed\tabularnewline
\hline 
false & - & function return\tabularnewline
\hline 
false & \textquotedbl{}`continue\textquotedbl{} & next iteration of innermost loop\tabularnewline
\hline 
false & \textquotedbl{}`break\textquotedbl{} & termination of innermost loop\tabularnewline
\hline 
false & (other) & propagation of an error from the command\tabularnewline
\hline 
\end{tabular}
\par\end{centering}

\caption{The result of command and error code effects used for internally build-in
commands. Note, user can expand the language error codes as she wishes
adding new functionality to the language. The special error codes
(that are not to be displayed for the user or caught) should start
with character {}```''.\label{tab:The-error-codes}}

\end{table}

\subsection{Expressions interpretation modes}

In T2Script, each command definition must include information about
the mode of the expressions interpretation. This mode affects the
VM policy of handling expressions passed to the command with parameters
and is either:
\begin{description}
\item [{automatic}] in which VM handles all expressions virtually before
passing the execution flow to the command code;
\item [{on~demand}] in which application programmer is responsible for
manual handling%
\footnote{\emph{Manually} means, in this context, that programmer uses functions
and objects of \emph{VariableContext} type in the VM for handling
expressions.%
} of the expressions in desired data portions from parameters. In this
mode, expressions are passed to the command in \emph{raw} format (not
interpreted).
\end{description}
This approach enables application programmer of constructing eval\index{eval}-type
commands (like \noun{mechanize\index{mechanize@\noun{mechanize}}}
that is {}``eval command'' in T2Script).

\subsection{Internal T2Script commands\label{sub:Internal-T2Script-commands}}

T2Sctipt defines some basic commands that are used as a language control,
variables, functions and events handling, manipulation of arrays and
timers, interpretation of expressions strings and numbers manipulation,
loading scripts modules, invoking other processes and metaprogramming.
The most significant commands that are available initially in T2Script
are listed in table \ref{tab:The-most-significant}. %
\begin{table}[h]
\begin{centering}
\begin{tabular}{|c|c|c|}
\hline 
Command & Parameters & Blocks\tabularnewline
\hline 
\hline 
\noun{if} & 1 & 2\tabularnewline
\hline 
\noun{repeat} & 1 & 1\tabularnewline
\hline 
\noun{while} & 1 & 2\tabularnewline
\hline 
\noun{foreach} & 3 & 1\tabularnewline
\hline 
\noun{for} & 3 & 2\tabularnewline
\hline 
\noun{break} & - & -\tabularnewline
\hline 
\noun{continue} & - & -\tabularnewline
\hline 
\noun{throw} & 1 & -\tabularnewline
\hline 
\noun{catch} & 1 & 1\tabularnewline
\hline 
\noun{mlc} & 1 & -\tabularnewline
\hline 
\noun{mlcext} & 2 & -\tabularnewline
\hline 
\noun{mechanize} & 1 & -\tabularnewline
\hline 
\noun{load} & 2 & -\tabularnewline
\hline 
\noun{runscript} & 2 & -\tabularnewline
\hline 
\noun{envrs} & 3 & -\tabularnewline
\hline 
\noun{expr} & 1 & -\tabularnewline
\hline 
\noun{settimer} & 3 & 1\tabularnewline
\hline 
\end{tabular}\begin{tabular}{|c|c|c|}
\hline 
Command & Parameters & Blocks\tabularnewline
\hline 
\hline 
\noun{function} & (no limit) & -\tabularnewline
\hline 
\noun{put} & 3 & -\tabularnewline
\hline 
\noun{functiondel} & 1 & -\tabularnewline
\hline 
\noun{return} & 1 & -\tabularnewline
\hline 
\noun{result} & 1 & -\tabularnewline
\hline 
\noun{args} & (no limit) & -\tabularnewline
\hline 
\noun{trigger} & 1 & -\tabularnewline
\hline 
\noun{null} & - & -\tabularnewline
\hline 
\noun{setvar} & 2 & -\tabularnewline
\hline 
\noun{delvar} & (no limit) & -\tabularnewline
\hline 
\noun{isset} & 2 & -\tabularnewline
\hline 
\noun{isnumeric} & 2 & -\tabularnewline
\hline 
\noun{setarray} & 3 & -\tabularnewline
\hline 
\noun{delarray} & 1 & -\tabularnewline
\hline 
\noun{arraysize} & 2 & -\tabularnewline
\hline 
\noun{isarray} & 2 & -\tabularnewline
\hline 
\noun{runfile} & 1 & -\tabularnewline
\hline 
\end{tabular}
\par\end{centering}

\caption{The most significant internally predefined T2Script language commands\index{command}.\label{tab:The-most-significant}}

\end{table}

The application programmer or an application in which the T2Script
module is embedded is able to disable or add selected commands (i.e.
language structures) during configuration stage and at run\nobreakdash-time.

We look upon some of the internal commands as constructs not present
in other programming languages and therefore not intuitive for the
user at a very beginning. The constructs follow strictly the format
of commands presented in section \ref{sub:Format-of-commands}.

\subsubsection{\noun{While}}

In T2Script\noun{, while\index{while@\noun{while}}} command is a
standard \emph{while loop}\index{loop} with exception it accepts
2 blocks separated with \noun{else} from which second block executes
once if and only if the loop condition is \emph{false} at the moment
when the program flow enters the condition. The first block is standard
loop's body that executes each time the condition is \emph{true}.
Listing \ref{lis:While-loop.} presents the example of \noun{while}
command usage. The \noun{textout\index{textout@\noun{textout}}} command
is used in the example for outputting the parameter to standard output.
The loop never executes \noun{null\index{null@\noun{null}}} command
that is used here as a placeholder because of the \emph{false} control
condition.

\begin{lstlisting}[caption={\noun{While} loop. },{float=h},label={lis:While-loop.}]
while $_false {
	null;
	// this code is never executed
} else {
	textout This code is executed only once;
}
\end{lstlisting}

\subsubsection{\noun{For}}

\noun{For} command is T2Script version of \emph{for loop\index{loop}}
from C++ language. \noun{For\index{for@\noun{for}}} accepts 2 blocks
separated by \noun{every\index{every@\noun{every}}} and 3 parameters,
being the most complex internal T2Script command. First block is standard
loop's body and second block is executed always after each entered
loop's iteration (even if loop flow was altered with \noun{continue\index{continue@\noun{continue}}}
or \noun{break\index{break@\noun{break}} }commands\noun{)}.

The parameters to the command are:
\begin{enumerate}
\item a name of the variable;
\item an initial value set to variable;
\item a control condition of the loop.
\end{enumerate}
The example loop is shown in listing \ref{lis:For-loop.}.

\begin{lstlisting}[caption={\noun{For} loop. },{float=h},label={lis:For-loop.}]
for i 0 $?[< $i 10] {
	if $?[eq $i 5] {
		continue;
	}
	textout $i;
} every {
	inc i;
}
\end{lstlisting}

Before loop starts, the variable \emph{i} is set to \emph{0}. Then,
the loop iterates 10 times until \emph{i} value reaches \emph{10}.
The value of \emph{i }is displayed on every iteration except when
the value of \emph{i} is equal to \emph{5}. After each loop iteration
(including the one when \emph{i} is equal to \emph{5})\noun{, inc\index{inc@\noun{inc}}}
command increments the value of the variable by \emph{1}. After the
control condition becomes \emph{false }and loop finishes, variable
\emph{i} still exists with value \emph{10}. The presented example
uses \emph{expressions} that are described in section \ref{sec:Expressions}.

\subsubsection{Eval-type commands: \noun{Mechanize} and \noun{MLC}}

Basic eval-type command in T2Script is \noun{mechanize\index{mechanize@\noun{mechanize}}.
}The command\noun{ }evaluates\noun{ }the other command provided in
the parameter. Listing \ref{lis:Mechanize-command-example.} shows
the example of usage of \noun{mechanize}. 
\begin{lstlisting}[caption={\noun{Mechanize} command example.},{float=h},label={lis:Mechanize-command-example.}]
setvar prog setvar name;
mechanize $prog. John;
\end{lstlisting}

In the result of this program, variable \emph{name }is set to value
{}``John''.

The more complex version of \noun{mechanize} is \noun{mlc\index{mlc@\noun{mlc}}}
(\emph{multi-line command}) that evaluates multiple commands provided
in the parameter and separated by separator (by default {}``||'').
User chooses\noun{ mlc} command when she inputs only one-line script
program. Listing \ref{lis:MLC-command-example.} shows the example
of usage of \noun{mlc}.

\begin{lstlisting}[caption={\noun{mlc} command example.},{float=h},label={lis:MLC-command-example.}]
setvar i -10;
while $?[< $i 0]. mlc textout $i||inc i;
\end{lstlisting}

The provided example uses \noun{mlc} to evaluate 2 (two) commands
\noun{textout} and \noun{inc}. Note, \noun{while} is an in\nobreakdash-line
version of \noun{while }that doesn't accept blocks and executes only
one instruction provided with second parameter. T2Script provides
also in\nobreakdash-line versions of commands: \noun{if}, \noun{repeat,
while }and\noun{ foreach}. These versions are designed to be used
with \noun{mlc} commands.

\section{Expressions\label{sec:Expressions}}

Expressions\index{expression}\nomenclature{Expression}{Structure that evaluates meaningful language terms.}
are used to evaluate meaningful language terms. The expressions in
T2Script fall into four categories: \emph{variables} (see section
\ref{sub:Variables}), \emph{constants} (see section \ref{sub:Constants}),
\emph{complex expressions} (see section \ref{sub:Complex-expressions})
and \emph{function calls} (see section \ref{sub:Functions-calling}).

\subsection{Format of expressions\label{sub:Format-of-expressions}}

User injects expressions into commands parameters to update them with
desired results of expressions. In the language, the user-data (including
strings and numbers) is not delimited with any characters and therefore
is considered of greater importance then expressions%
\footnote{In many known programming languages, the user data is delimited, e.g.
using quotes and expressions are not delimited.%
}. The concept of great importance of user-data derives from the language
purpose (string processing and command processing). Consequently,
the expressions must be delimited in order to separate them from the
user-data.

\begin{figure}[h]
\begin{centering}
\includegraphics[width=0.8\textwidth]{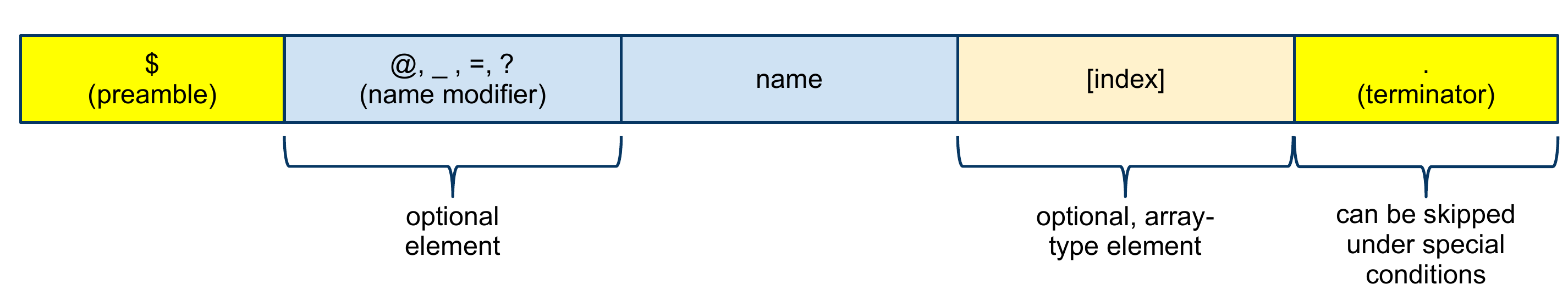}
\par\end{centering}

\caption{Structure of generic expression in T2Script.\label{fig:Structure-of-expression}
Identical colors of elements express similarity. The expression is
delimited with preamble (dollar sign) and terminator (dot). If used,
name modifier is one of the following symbols: @, \_, =, ?. Array-type
expressions have index between square brackets.}

\end{figure}

The structure of the expression in T2Script is presented in fig. \ref{fig:Structure-of-expression}.

We can distinguish few elements of expression:
\begin{description}
\item [{preamble}] (dollar sign) introduces the start of the expression;
\item [{name~modifier}] ({}``@'', {}``\_'', {}``='' or {}``?'')
determines a type of the expression;
\item [{name}] (consists of arbitrary number of Unicode characters except
square brackets {}``{[}{}``, {}``{]}'' and {}``.''; first character
must not be one of name modifiers characters; can contain parenthesis
{}``({}`` and {}``)'' but if and only if the opening elements
number are equal to closing elements number) is used to identify expression;
\item [{index}] (between square brackets) is an extension of the expression
(it can recursively contain another expression);
\item [{terminator}] (dot) is the last character of the expression and
is used to concatenate expression with following user-data or expression.
\end{description}
The terminator\index{expression terminator} element can be skipped
under special conditions if either:
\begin{enumerate}
\item the expression is the last element of the last parameter of the command
(no user-data or expression follows it)

\begin{enumerate}
\item includes structure where the expression is the last element before
block beginning character in block-command;
\end{enumerate}
\item the expression is the last element inside the index of another expression;
\item the expression is the last or unique element of any argument of an
operator in complex expressions.
\end{enumerate}
If the user is not certain whether she is able to skip terminator
character she safely does not skip it.

\subsection{Variables\label{sub:Variables}}

A \emph{variable} expression is recognized by having name modifier
equal to {}``@'' or not having name modifier. T2Script is untyped
language with dynamic type checking. Thus, user doesn't specify variables
type explicitly and one variable is able to carry any type of data
(value of the variable). The type is checked at run-time when the
value of the variable\index{variable}\nomenclature{Variable}{Variable is type of expression without name modifier or with name modifier "@" that stores user defined value of any type.}
is used with specific commands. During its life-time, one variable
can change the type of data it contains.

Variable is either global or local. Global variables don't have any
name modifiers and are visible in all programs scopes. Local variable
has name modifier equal to {}``@'' and is visible in a scope of
a function it has been set. Local variable persists during function
run and then it is automatically removed from the VM's memory.

Additionally, T2Script supports one-dimensional arrays variables.
This type of variables include index after the name. Size of the arrays
doesn't need to be given when the user sets an array and this size
may change at runtime. Index must be numeric for some of the commands
operate on arrays but T2Script has also limited support (i.e. support
for manual operations on them) for associative arrays with indexes
that are not numeric.

Different types of variables are presented in listing \ref{lis:Different-types-variables}.

\begin{lstlisting}[caption={Different types of variables in T2Script. },{float=h},label={lis:Different-types-variables}]
// global variable's value
textout $account_number;

// local variable's value
textout $@name;

// global array variable's value
textout $sinus[90];

// global array variable's value with index of local array variable's value
textout $sinus[$@angle[5]];

// associative global array variable's value
textout $birthday[Piotr];
\end{lstlisting}

\subsection{Constants\label{sub:Constants}}

A \emph{constant}\index{constant} is type of expression\nomenclature{Constant}{Constant is type of expression with underscore character as name modifier that stores value of any type that is not possible to modify by the user.}
that is recognized by having name modifier equal to {}``\_'' (underscore).
T2Script has build-in set of constants (see table \ref{tab:Build-in-T2Script-constants}).
Additionally, application programmer extends this set by defining
constants in her application. User is not able to modify constants
directly (in opposite to variables) although she can use their current
values.%
\begin{table}[h]
\begin{centering}
\begin{tabular}{|c|c|}
\hline 
Constant & Value description\tabularnewline
\hline 
\hline 
\$\_true & true value (zero value)\tabularnewline
\hline 
\$\_false & false value (non-zero value)\tabularnewline
\hline 
\$\_empty & empty value\tabularnewline
\hline 
\$\_parent\_name & name of parent command\tabularnewline
\hline 
\$\_parent\_param & parameters of parent command\tabularnewline
\hline 
\$\_owner\_name & name of the current command\tabularnewline
\hline 
\$\_owner\_param & parameters of the current command\tabularnewline
\hline 
\$\_time & current time (in local format)\tabularnewline
\hline 
\$\_date & current date (in local format)\tabularnewline
\hline 
\$\_Pi & Pi value\tabularnewline
\hline 
\$\_\textbackslash{}n, \$\_\textbackslash{}r\textbackslash{}n and
\$\_\textbackslash{}r & end of line markers\tabularnewline
\hline 
\$\_\textbackslash{}t & horizontal tab\tabularnewline
\hline 
\$\_\textbackslash{}\$ & dollar sign\tabularnewline
\hline 
\$\_\textbackslash{}s & space\tabularnewline
\hline 
\$\_lparen, \$\_rparen & {}``({}`` and {}``)''\tabularnewline
\hline 
\$\_ltabparen, \$\_rtabparen & {}``{[}{}`` and {}``{]}''\tabularnewline
\hline 
\$\_lcurlparen, \$\_rcurlparen & {}``\{'' and {}``\}''\tabularnewline
\hline 
\$\_\textbackslash{}u(\emph{val}) & Unicode character of decimal value \emph{val}\tabularnewline
\hline 
\end{tabular}
\par\end{centering}

\caption{Build-in T2Script constants and their values descriptions.\label{tab:Build-in-T2Script-constants}
In T2Script, \emph{false} value is represented by 0 (zero character),
\emph{true} value is represented by anything else (including empty
string). Parent command is a command that runs current command or
is a block-command in which block the current command runs (if command
doesn't have a parent, parrent-type constants show information of
current command).}

\end{table}

Note, application programmer assigns a string or a function to T2Script
constant. Consequently, if function was assigned to a constant, the
value of constant may change for subsequent calls. \emph{Constancy\index{constancy}}
is thus expressed as lack of ability to change the constant value
from the script level, not as value's permanence.

\subsection{Complex expressions\label{sub:Complex-expressions}}

\begin{figure}[h]
\begin{centering}
\includegraphics[width=0.7\textwidth]{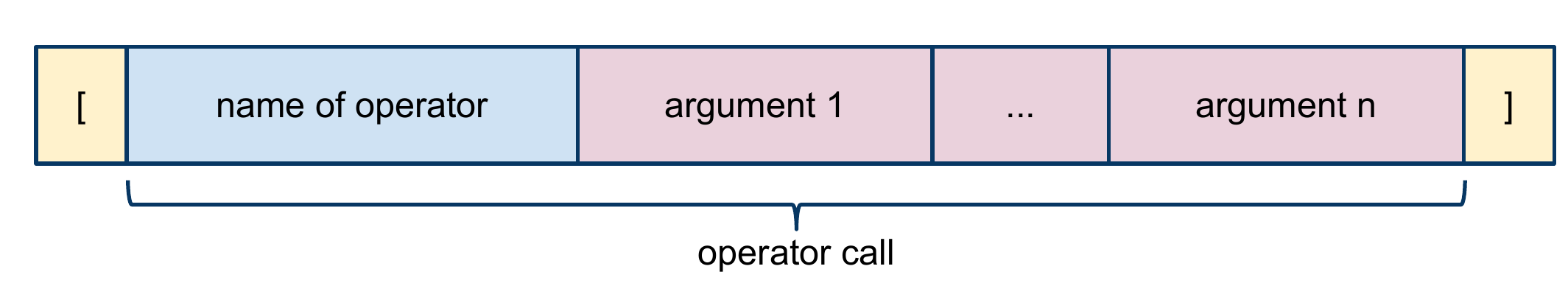}
\par\end{centering}

\caption{Structure of an index of a generic complex expression\label{fig:Structure-of-index}.
Identical colors of elements express similarity. Each element in operator
call is separated by at least one space. Argument of an operator may
recursively be another operator call if placed in between parentheses.}

\end{figure}

A\emph{ complex expression}\index{complex expression}\nomenclature{Complex expression}{Complex expression is a type of expression that has name modifier equal to question mark character and index which contains operators and their arguments. Complex expressions are often used to evaluate mathematical expressions.}
is an expression that has name modifier equal to {}``?'' (question
mark) and index which contains \emph{operators} and their \emph{arguments}.
Therefore, the main part of complex expression is its index which
contains the term to evaluate. The structure of the index of complex
expression is presented in fig. \ref{fig:Structure-of-index}. %

Complex expressions call operators and pass them arguments. User uses
complex expressions e.g. to evaluate mathematical expressions. The
syntax of operators in T2Script is similar to function calls used
in functional languages. T2Script however doesn't support lazy evaluation
of expressions. Moreover, all the expressions in the complex expression
are evaluated before returning the final result. Example of complex
expressions usage in T2Script are presented in listing \ref{lis:Examples-complex-expressions}.
\begin{lstlisting}[caption={Examples of usage of complex expressions.},{float=h},label={lis:Examples-complex-expressions}]
// outputs "101"
textout $?[+ 2 (- 7 8) 100];

// true only if @name value is equal to "Piotr" or "John"
if $?[or (eq $@name Piotr) (eq $@name John)] {
    textout Name correct;
}
\end{lstlisting}

\subsubsection{Operators\label{sub:Operators}}

Operators are functions (functions are described in section \ref{sec:Functions})
to use with complex expressions\nomenclature{Operator}{Operator is function type to use with complex expressions.}.
T2Script provides build-in operators\index{operators} that are ready
to be used with complex expressions. The operators are divided in
8 categories:
\begin{enumerate}
\item Basic arithmetic operators. Include operators for integers and floating
point numbers.
\item Other arithmetic operators. Include modulo, power, square root, logarithms,
minimum, etc.
\item Bitwise operators.
\item Rounding routines operators used for rounding floating point numbers.
\item Logical operators designed for control conditions.
\item Relational operators used to compare strings and numbers and for regular
expressions (refer to section \ref{sec:Regular-expressions}).
\item Strings operators used to handle strings operations.
\item Other advanced operators. Include variable assignments, existential
operator, command execution operator.
\end{enumerate}
The full list of build-in operators with corresponding categories
is presented in table \ref{tab:List-of-operators}.%
\begin{table}[h]
\begin{centering}
\begin{tabular}{|c|c|}
\hline 
Category & Operators\tabularnewline
\hline 
\hline 
1. & +, -, {*}, /, +., -., {*}., /.,\tabularnewline
\hline 
2. & \%, {*}{*}, sqrt, ln, logn, exp, abs, min, max, tohex\tabularnewline
\hline 
3. & \textasciitilde{}, \&, |, \textasciicircum{}, <\textcompwordmark{}<,
>\textcompwordmark{}>\tabularnewline
\hline 
4. & round, roundto, ceil, floor\tabularnewline
\hline 
5. & ! (not), ?| (or), ?\& (and)\tabularnewline
\hline 
6. & == (eq), != (ne), <= (le), >= (ge), < (lt), > (gt), :== (eqic), :!=
(neic), =\textasciitilde{}, =\textasciitilde{}\textasciitilde{}, comp\tabularnewline
\hline 
7. & :+ (concat), empty?, len, num?, float?, substr, strpos, strposic,
word, char, upcase, downcase\tabularnewline
\hline 
8. & =, exists?, !!, @@, ??\tabularnewline
\hline 
\end{tabular}
\par\end{centering}

\caption{List of build-in operators.\label{tab:List-of-operators} Operators
are separated with coma in the table. If operator has alternative
names, they are shown in parentheses.}

\end{table}

If user uses operator that is not build-in operator, and if function
exists with the name of the used operator, it is called. Thus, user
defines own operators functions in the script and calls them with
complex expression structure (see section \ref{sub:Functions-calling}).

\section{Functions\label{sec:Functions}}

\begin{figure}[h]
\begin{centering}
\includegraphics[width=0.7\textwidth]{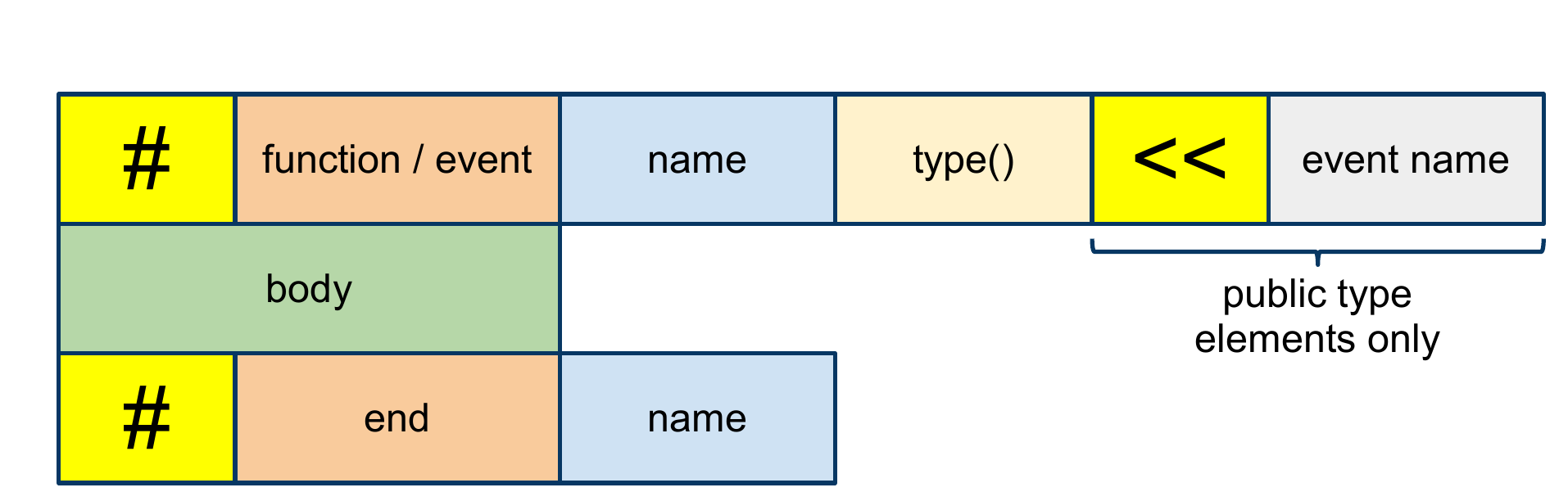}
\par\end{centering}

\caption{Generic structure for definitions of functions and events.\label{fig:Generic-function}
Identical colors of elements express similarity. Some elements (including
event name) are only allowed in functions if the type \emph{public()}
is chosen.}

\end{figure}

Functions\index{function}\nomenclature{Function}{Function is high level object that contains commands. Each function has a name and a type.}
in T2Script are high level objects identified by a name. Function
is composed of commands. T2Script shares common structure for definitions
of functions and events. The structure is presented in fig. \ref{fig:Generic-function}.

The function name must be unique for the script file otherwise redefinition
error will be signaled to the user. Lines starting with hash must
not have semicolon\index{semicolon} in the end. User chooses the
type for the function definition. The type allowed for functions is
one of the following types:
\begin{description}
\item [{private()}] describes function with no event assigned to it;\index{private()}
\item [{public()}] describes function with one event assigned to it;\index{public()}
\item [{operator()}] is identical to \emph{private()} but describes the
function that is used as an operator in complex expressions%
\footnote{It is not an obligation to use \emph{operator()} type. Essentially,
every function can be used as an operator. It is, however, good practice
to use it. The script is easier to read.%
}.\index{operator()}
\end{description}
If the function type is \emph{public(),} the existing event name must
be provided in order to assign it to the function. In the result,
event generator calls the function when the event occurs.

\subsection{Functions arguments}

Function accepts arbitrary number of arguments\index{arguments}\nomenclature{Argument}{Argument is a parameter of function.}.
Like in variables, in arguments the type is not specified. When user
calls the function, unnamed arguments are available in the function
in local array variable \emph{@arg}. The first argument has index
equal to \emph{0}. When the event generator calls the function, arguments
are copied from the event and are usually named accordingly to the
content. User may also use \noun{args} command and name unnamed arguments.
The \noun{args} command is used only once in the function (usually
as the first command). \noun{args} command also enforces minimal number
of arguments that must be passed to the function. If the number of
arguments passed is smaller than the number expected by \noun{args},
the error is thrown. Example code presenting unnamed and named arguments
is presented in listing \ref{lis:Unnamed-and-named}.

\begin{lstlisting}[caption={Unnamed and named arguments in functions. },{float=h},label={lis:Unnamed-and-named}]
// function uses unnamed arguments
#function unnamed private()
	textout First name: $@arg[0];
	textout Last name: $@arg[1];
#end unnamed

// function uses named arguments
#function named private()
	args first last;
	textout First name: $@first;
	textout Last name: $@last;
#end named

// function uses event's arguments
#function new_user public() << on_new_user
	textout Welcome $@username;
#end new_user
\end{lstlisting}

Local arguments to the function may also be \emph{accumulated} on
the function stack before a call to the function with \noun{put\index{put@\noun{put}}}
command. This type of accumulated arguments may have arbitrary names.
The accumulated arguments are available in the function only in one
subsequent call and then are removed with other local variables when
function terminates.

\subsection{Functions result}

Each function returns a result. User alters the default result of
function that is equal to \emph{\$\_empty} (no value) with the following
commands:
\begin{description}
\item [{\noun{return}}] sets the result and immediately returns it changing
flow of the program;\index{return@\noun{return}}
\item [{\noun{result}}] sets the result that will be returned when the
function terminates without changing flow of the program.\index{result@\noun{result}}
\end{description}

\subsection{Functions calls\label{sub:Functions-calling}}

User calls a function to pass the arguments (if any passed) and start
execution of function code. There are two modes of function calls:
\begin{enumerate}
\item command\nobreakdash-mode call when arguments are corresponding to
words in a given argument\nobreakdash-string (one argument is one
space\nobreakdash-separated word);
\item call when arguments are not corresponding to words (one argument may
be more than one space\nobreakdash-separated word).
\end{enumerate}
The first mode is used with command \noun{function\index{function@\noun{function}}}
and expressions\index{expression} with name modifier equal to {}``=''
and arguments in an index. In this calls, passed argument is in fact
one string that is split into words based on spaces -- subsequent
words are accommodated with subsequent indexes of \emph{@arg} array.

The second type is used with calls of operators -- the arguments may
contain many words or empty strings. In the example in listing \ref{lis:Different-types-of},
we show the function that returns only first argument passed and is
called in different ways with string argument {}``first second''
returning different results based on the call type.

\begin{lstlisting}[caption={Different types of the functions calls. },{float=h},label={lis:Different-types-of}]
#function fnc private()
	return $@arg[0];
#end fnc

#function test public() << on_load
	// outputs "first" (type 1)
	textout $=fnc[first second];
	
	// outputs "first" (type 1)
	function fnc first second;
	
	// outputs "first second" (type 2)
	textout $?[fnc (concat first $_\s second)];
#end test
\end{lstlisting}

\section{Events\label{sec:Events}}

User creates own events that she exposes to other scripts and other
users. The events in T2Script are based on observer pattern:
\begin{itemize}
\item the user creates an event with unique name;
\item public function registers itself to be called with an event;
\item an event is triggered.
\end{itemize}
Events\index{event}\nomenclature{Event}{Event is an action that triggers registered functions. Events work based on observer pattern. Events give to the user a possibility of  exposing own events to other users from the script.}
definitions are optional in the language. Application programmer is
able to disable them for security reasons. User is not able to trigger
build-in events that application programmer defined in the application
(using build\nobreakdash-in functions of VM).

The event definition is presented in fig. \ref{fig:Generic-function}.
All the naming conditions used for functions (except types) also hold
for events. The possible types of events are:
\begin{description}
\item [{single()}] when triggered, calls only one, non\nobreakdash-deterministically
chosen function from the set of registered functions for this event;\index{single()}
\item [{multi()}] when triggered, calls all of the functions from the set
of registered functions for this event.\index{multi()}
\end{description}

\subsection{Events variables\label{sub:Events-variables}}

An event structure is very similar to normal function structure. It
has local variables. These variables in the moment of triggering are
copied to the called registered function(s) with the corresponding
names. This mechanism enables user to create events variables. The
example of event definition with one event variable \emph{@username}
is presented in listing \ref{lis:Example-of-event}.

\begin{lstlisting}[caption={Example of event definition.  },{float=h},label={lis:Example-of-event}]
#event on_new_user multi()
	args username;
	trigger;
#end on_new_user

#function create_user private()
	// here creation of new user "Piotr"
	exp $=on_new_user[Piotr];
#end create_user
\end{lstlisting}

\subsection{Events results}

Events return values like other functions. User also checks a result
of registered functions by supplying a parameter to \noun{trigger\index{trigger@\noun{trigger}}}
command. The parameter becomes a name of local array with values of
the results of the registered function(s). In case the event has \emph{single()}\index{single()}
type, the array contains only one element. In case there are no registered
functions for the event, the array is not set. Example of event giving
an approval for an action is presented in listing \ref{lis:Example-of-event-with-result}.

\begin{lstlisting}[caption={Example of an event returning a result.},{float=h},label={lis:Example-of-event-with-result}]
#event on_approval multi()
	args action;
	// default result
	result $_true;
	trigger @votes;
	foreach @vote in @votes {
		if $?[! $@vote] {
			return $_false;
		}
	}
#end on_approval

#function block_shutdown public() << on_approval
	whitelist @action shutdown exit;
	if $processing {
		return $_false;
	}
#end block_shutdown
\end{lstlisting}

The presented example first sets the result to \emph{\$\_true}, then
triggers an event and calls all registered functions. The array \emph{@votes}
is used to store the results from the functions. If any of the registered
function returns \emph{\$\_false}, then the event does not give an
approval for an action (returns \emph{\$\_false}). Function \emph{block\_shutdown
}is an example of registered function. It checks two actions: {}``shutdown''
and {}``exit'' with \noun{whitelist\index{whitelist@\noun{whitelist}}}
command and if program is processing while one of these actions is
triggered, it returns \emph{\$\_false}.

\section{Timers}

Timers\index{timer} are parts of program that repeat the execution
automatically based on given time interval.\nomenclature{Timer}{Timer is a part of program that repeat the execution automatically based on time interval.}
Timer is recognized by its name. If user sets a name of the timer
to {}``auto'', the VM generates timer name automatically. Timers
operations are handled with few timers\nobreakdash-related commands
including block command \noun{settimer}\index{settimer@\noun{settimer}}.
\noun{settimer} creates and runs the timer immediately. The example
of timer is presented in the listing \ref{lis:Timer-example.}.

\begin{lstlisting}[caption={Timer example.},{float=h},label={lis:Timer-example.}]
#function counter private()
	setvar @local Local variable;
	settimer auto 1000 10 {
		textout $@local;
	}
	setvar @local Hello;
#end counter
\end{lstlisting}

Timer copies local variables (from function or event) to its local
scope in the moment of creation. Therefore, in the example, a \emph{@local}
variable value change after creation of the timer (automatic name,
1000 millisecond interval and 10 iterations) is not affecting the
value of \emph{@local} in the timer. The mechanism of copying the
variables is similar to copying mechanism used in events when calling
registered functions (see section \ref{sub:Events-variables}).

\part{Advanced programming techniques}

\section{Exceptions handling}

An \emph{exception}\index{exception} in T2Script is an error message
in the script\nomenclature{Exception}{Exception is an error message in the script.}.
When the exception propagates, a current program execution is stopped
and the exception is presented for the user. User may ignore the exception
by catching it. User catches the exception with command \noun{catch\index{catch@\noun{catch}}}.
If the parameter is present in \noun{catch} command, user saves the
exception to variable for error examination purposes from the script
level. Additionally, user may throw own exception from the variable
using \noun{throw} command. The example program using exceptions is
presented in listing \ref{lis:Exceptions-handling.}.

\begin{lstlisting}[caption={Exceptions handling.},{float=h},label={lis:Exceptions-handling.}]
#function fnc private()
	catch @err {
		setvar @msg This is error message;
		throw @msg;
	}
	textout $@err;
#end fnc
\end{lstlisting}

The presented example first throws an error and then catches it and
display it to the user an error message. The error propagation mechanism
presented in the example is intra\nobreakdash-functional but it is
valid also for inter\nobreakdash-functional scenarios -- the errors
propagate between the functions calls.

\section{Regular expressions\label{sec:Regular-expressions}}

T2Script includes extended support for regular expressions\index{regular expression}
with commands and operators. Regular expressions in T2Script are based
on \emph{Boost Regex Library} \cite{Karlsson2005}. The syntax of
the regular expressions is Perl syntax and follows \emph{Regex} manual%
\footnote{http://www.boost.org/doc/libs/1\_45\_0/libs/regex/doc/html/boost\_regex/syntax/perl\_syntax.html%
}. The regular expressions syntax underlies the T2Script syntax --
if a regular expressions uses reserved T2Script literals, these must
be escaped prior to passing them to regular expression.

\section{Metaprogramming\label{sec:Metaprogramming}}

\begin{figure}[h]
\begin{centering}
\includegraphics[width=0.7\textwidth]{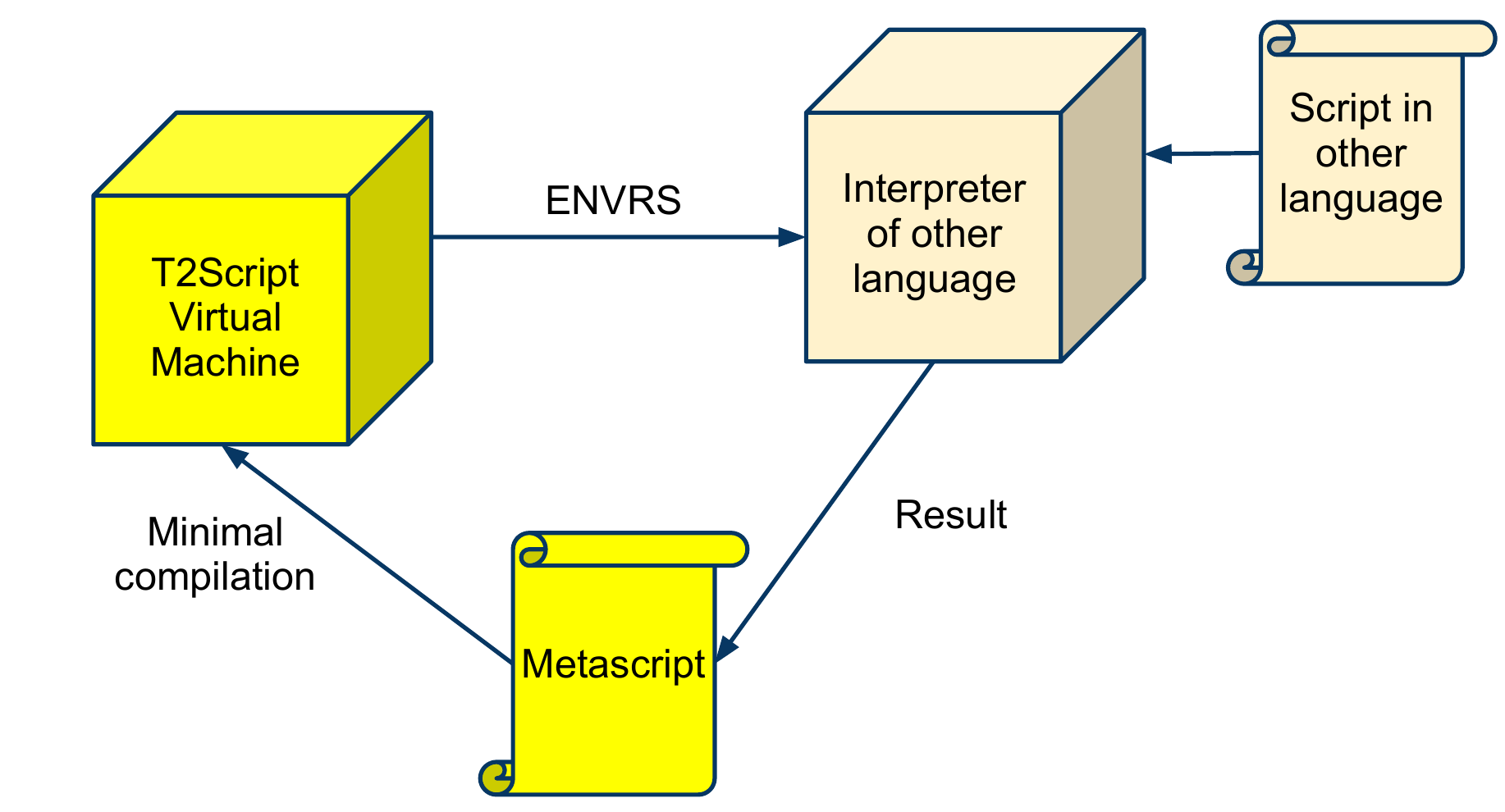}
\par\end{centering}

\caption{Metaprogramming technique with T2Script.\label{fig:Metaprogramming-technique-with}
The metascript does not compile in normal compilation process but
instead passes minimal\protect\nobreakdash-compilation that minimizes
the cost of using an external interpreter of other language by reducing
compilation time.}

\end{figure}

T2Script VM supports metaprogramming\index{metaprogramming} techniques
with command \noun{envrs\index{envrs@\noun{envrs}}. }The command
execution includes:
\begin{itemize}
\item execute any other interpreter with given parameters,
\item pipeline the result back to VM,
\item evaluate the result as T2Script.
\end{itemize}
The process of calling and pipelining is optimized and the additional
cost caused by using an external interpreter is minimized. T2Script
programs that are the result of this process do not pass through the
whole process of compilation but through the process called \emph{minimal\nobreakdash-compilation}.
Noticeably, scripts used with minimal\nobreakdash-compilation\index{minimal-compilation}
must not contain definitions of events or functions and are executed
with a default context (see section \ref{sub:Context-of-commands}).
The process is illustrated in the fig. \ref{fig:Metaprogramming-technique-with}. 

\section{Dynamic-Link Libraries (DLL) plugins\label{sec:Dynamic-Link-Libraries-(DLL)}}

T2Script has internal support for DLL\index{DLL} libraries including
loading and unloading DLL files as well as mapping DLL functions into
T2Script functions. The mapping coverts the local function variables
to C++ array and maps back the result of C++ function into T2Script
function's result using one command \noun{dlllocal}\index{dlllocal@\noun{dlllocal}}.
This mechanism allows user to write plugins to applications that include
T2Script module.

Additionally, the application programmer may allow user to write callback
plugins\index{plugins} using one of the mechanisms that is provided
by specific operation system where the application is deployed (e.g.
message queries, shared memory).

\section{Debugging and Testing}

T2Script VM offers API for attaching the debugger\index{debugger}
at \emph{command level}. Debugger provides command by command execution
without partial expression execution. The debugger offers also watch\nobreakdash-lists
of all functions, global variables and timers. Local variables are
evaluated during the debugging process inside a function. The T2Script
module does not offer any GUI for debugging, it must be provided by
application programmer.

Language does not offer any test suite but it can be integrated without
afford by application programmer.

\part{Summary}

\section{Solutions to the given problems}

After the description of the language, we show the solution to the
problems given in section \ref{sec:Motivation}.

\subsection{Interaction with the user}

In CED programming the user interacts with the language by commands
and scripts with events definitions and functions. Therefore, the
language increases the interaction with the users. It is reflected
in T2Script language construct of command that provides natural way
of expressing the name and parameters.

\subsection{Abstraction of an events dispatcher}

We would like our language to abstract the events and dispatcher on
the language level. Therefore the design of the language includes
the events definition and events\nobreakdash-registered functions
(see section \ref{sec:Events}). The events are implicit parts of
the language and user is encouraged to use them even for simple programs.

\subsection{Separation of the text and expressions}

T2Script is mainly designed to operate on string data. As it was shown
in section \ref{sub:Format-of-expressions} and description of expressions
format, the expressions are separated from the text, not the text
from expressions (difference of importance). This is an experimental
design that we chose for our string handling language.

\subsection{Other languages modules}

We would like our language to be capable to easy extend with modules
written in other programming languages. Moreover, we would like the
language to do this with optimal cost. The solution for this was presented
in sections \ref{sec:Metaprogramming} and \ref{sec:Dynamic-Link-Libraries-(DLL)}.
Built\nobreakdash-in T2Scripts commands allow user to apply metaprogramming
techniques using other languages interpreters and to write simple
DLL plugins. The minimal\nobreakdash-compilation process optimizes
the cost of metaprogramming scripts run times.

\subsection{Static code analysis}

By usage of CED Programming code security issue is solved without
static analysis of the application. The solution we propose promotes
enforcement of contract during the run-time in the VM.

The description of disabling the selected commands of T2Script language
is presented in section \ref{sub:Internal-T2Script-commands}. The
application is able to apply the contract at run\nobreakdash-time
by disabling specific VM commands prior to unknown script execution.
That reduces the risk of executing a potentially dangerous script.
Contract schema is however not included in T2Script so this problem
is not fully solved.

\subsection{Language extensions}

By the same mechanisms that are used to disable selected commands,
the new commands are added by application programmer. This enables
the language extensibility with possibilities of fast creation of
new language structures etc. Furthermore, the user also extends the
language in limited ways by defining own operators (see section \ref{sub:Operators}).

\section{CED programming}

Summarizing, we define CED programming as a paradigm in which programming
language:
\begin{enumerate}
\item gives user natural way of using commands that are part of the language;
\item incorporates events handling mechanisms.
\end{enumerate}
In CED programming, everything is based on command entity and events
are used to invoke groups of commands.

\section{Evaluation\label{sec:Evaluation}}

T2Script is dynamic untyped language (or scripting language). The
comparison of system and scripting programming languages reveals that
scripting languages shorten the development time from 4 to 8 times
in average and reduce number of lines from 100 to 1000 times in comparison
to system programming languages \cite{Ousterhout2002}. Thus T2Script
effectively reduces the workload needed to write scripts based on
commands and expressions. 

T2Script module is designed to be embedded into custom application
in 24 hours work period of experienced programmer. It is noticeably
short time in comparison to the time needed to implement own scripting
module. Unicode format of characters used in T2Script allows internationalization.

T2Script offers natural approach in handling application commands,
network and interfaces. Commands nowadays are used in some form in
almost all applications. The nature of commands however is very simple
and may an obstacle when implementing complex programs -- this is
why T2Script introduces block\nobreakdash-commands available from
script files. The T2Script command parameters (except the last one)
correspond to words in the user\nobreakdash-string. This simple approach
might be an advantage or a disadvantage dependent on the application.

T2Script commands are relatively easy to understand in comparison
to expressions that require more attention from a user. User has wide
possibilities of writing modules containing events and later to expose
them to other users. It encourages the code reusability and modularity.

T2Script offers many extensions possibilities using metaprogramming
techniques or DLL plugins and therefore gives the user a powerful
tool while relieving application programmer.

\section{Future work}

In this document, we mentioned few times how application programmer
integrates T2Script module into custom application. This topic however
should be extended in the next publications to be discussed in more
details. 

We can see the possibility of future work in investigation of the
syntax and performance measures of T2Script in comparison to other
programming languages.

Another possible direction is implementing a test suite for the users
in T2Script module. The test suite can adopt existing commands and
VM functions used now for debugging. It will however require tests
at different levels and include an event simulator.

As the T2Script files are compiled to a bytecode, this bytecode format
can be written to a file. Currently the compilation occurs internally
but, in the future, it may be possible to give the user an opportunity
to save the scripts files in bytecode format.

Another possible future extension is to build a standardize processes
of contract enforcement and assignment for T2Script, so that the user
will trust the unknown scripts during the execution.

\pagebreak{}

\settowidth{\nomlabelwidth}{Application programmer}
\printnomenclature{}

\printindex{}

\bibliographystyle{plain}
\phantomsection\addcontentsline{toc}{section}{\refname}\bibliography{t2script_intro}

\end{document}